\documentclass[twocolumn,showpacs,superscriptaddress,preprintnumbers,amsmath,amssymb,prb]{revtex4}

\usepackage{graphicx}
\usepackage{dcolumn}
\usepackage{bm}

\begin{document}

\title{Spin Dependence of Interfacial Reflection Phase Shift at Cu/Co Interface }
\author{Yuan Xu}
\affiliation{State Key Laboratory for Surface Physics, Institute
of Physics, Chinese Academy of Sciences, P.O. Box 603, Beijing
100080, China}
\author{Xi Mi}
\affiliation{ Surface Physics Laboratory (National Key
Laboratory), Fudan University, Shanghai 200433, China }
\author{Y. Z. Wu}
\affiliation{ Surface Physics Laboratory (National Key
Laboratory), Fudan University, Shanghai 200433, China }
\author{Ke Xia}
\affiliation{State Key Laboratory for Surface Physics, Institute
of Physics, Chinese Academy of Sciences, P.O. Box 603, Beijing
100080, China}

\date{\today}

\begin{abstract}
The spin dependent reflection at the interface is the key element
to understand the spin transport. By completely solving the
scattering problem based on first principles method, we obtained
the spin resolved reflectivity spectra. The comparison of our
theoretical results with experiment is good in a large energy
scale from Fermi level to energy above vacuum level. It is found
that interfacial distortion is crucial for understanding the spin
dependence of the phase gain at the Cu$|$Co interface. Near the
Fermi level, image state plays an important role to the phase
accumulation in the copper film.

\end{abstract}


\maketitle

\section{Introduction}
The spin dependent transparency of the nonmagnetic
metal$|$ferromagnet (NM$|$FM) interface plays a crucial role in
the magnetoelectronics. The spin dependent transmission and
reflection result in the spin dependent interface resistance,
which dominates the giant magneto-resistance effect\cite{GMR}.
More interestingly, the spin dependent reflection at the NM$|$FM
interface also yields the spin torques\cite{torques,
Brataas_circuit} at the interface. Spin torque across NM$|$FM can
be formulated in terms of the mixing
conductance\cite{Brataas_circuit}
$G_{mix}\equiv\frac{e^{2}}{h}[M-\sum_{nm}|r_{\uparrow}^{nm}||r_{\downarrow}^{nm}|exp(i(\phi_{\uparrow}^{nm}-\phi_{\downarrow}^{nm}))]$
, where $M$ is the total number of channels (eigenstates), and
$|r_{\uparrow(\downarrow)}^{nm}|exp(i\phi_{\uparrow(\downarrow)}^{nm})$
denotes the reflection from channel $m$ to channel $n$. The
reflection phase gain difference (PGD) defined as
$\phi_{\uparrow}^{nm}-\phi_{\downarrow}^{nm}$ plays an important
role to understand the spin torque.

However, measuring spin dependent transport through FM$|$NM is far
from trivial, since the electrons with different spin orientation
flow together in most of the electric transport process. It is
possible to probe the spin dependent transmission by making use of
the two current model\cite{fert, MSU}. The Andreev reflection (AR)
spectroscopy at ferromagnet-superconductor interfaces in ballistic
point contacts (PCAR)\cite{deJong95,PCAR}, as an alternative way
to measure the spin polarization, contains all the information of
the scattering matrix at the interface but experimental data are
very difficult to be resolved\cite{kexiaPRL}. It is deserved to
have one method which can measure the transmission or reflection
through an interface for different spins \emph{separately}.

Recently, Wu {\it et al.},\cite{Co_Cu_YZWu} have performed the
measurement of the spin-dependent electron reflection on the
Vacuum(Vac)$|$Cu$|$Co(001) film. Spin polarized incident electrons
were employed to detect spin dependent reflection at Cu$|$Co
interface. The PGD between two spin channels at Cu$|$Co interface
was obtained by measuring the relative shift of reflectivity
spectra for different spin polarization direction. Such an
experiment provides us an opportunity to test the theoretical
calculation of electronic transport with experiment in great
details, especially the subtle reflection phase. Furthermore, the
effect of atomic structure of interface on the spin transport can
be addressed.

In this paper, the phase accumulation and PGD in
Vac$|$Cu$|$Co(001) system is obtained from first principles
calculation and compared with experiment in a large energy scale.
The interfacial distortion is found to be important for
understanding the spin dependence of phase gain at Cu$|$Co
interface. Effects due to the distortion on the spin transport are
also discussed. Besides, we also estimate the possible phase gains
from image states at the surface of Cu film. Such extra phase gain
could be obvious near Fermi level.

This paper is organized as following: In Sec.II, we give a brief
introduction of our theoretical method. The modelling of
interfacial relaxation and the optimized structure are specified.
The experiment details will also be given in this section. In
Sec.III, we compare the theoretical results with experimental
data, where interfacial alloy and image potential are taken into
account. We also discuss the effect of structure relaxation on the
spin transport at interface. Our results will be summarized in
Sec.IV.

\section{Theoretical Method and Experiment}
Several works ~\cite{Fe_Ag_ab_initio_phase, Cu_Co_ab_initio_phase}
tried to calculate the phase gain at the NM$|$FM interface, but
the definition of the phase gain at a single interface for a given
energy is no unified due to the arbitrary phase when solving the
Schrodinger equation. Wei and Chou~\cite{Fe_Ag_ab_initio_phase}
obtained the phase gain at Ag$|$Fe interface by comparing quantum
well (QW) states positions of free standing Ag film and Ag film on
top of Fe. While An {\it et al.,}~\cite{Cu_Co_ab_initio_phase}
defined the phase gain at Cu$|$Co interface by a special fitting
wave function. In fact, what can be measured in experiment is not
the phase gain at single interface, but the sum of the phase gains
at two opposite interfaces or the phase gain difference of two
spin channels at the same interface.

Based on a recently developed first principles
method\cite{master}, we obtained the phase gains by calculating
the scattering matrix at interfaces. The arbitrary phase can be
removed by noticing that the incoming wave of one interface is the
outgoing wave of another interface for electrons scattering
between two opposite interfaces. The phase accumulation obtained
by the present method can be used to reproduce the positions of QW
states from the electronic structure.

In this method, the bulk and the interface potentials are
determined in the framework of the tight binding (TB) linearized
muffin tin orbital (MTO) with surface green function (SGF) based
on density functional theory(DFT) and local density
approximation(LDA). Disordered
systems\cite{Disorder_roughness_TMR} are treated using the layered
coherent potential approximation (CPA). As SGF is incorporated
into TB-LMTO, the electronic structure of interfaces and other
layered system can be treated in our framework. The
potentials\cite{spdf} obtained in this way were used as input to a
TB-MTO wave-function-matching (WFM) calculation of the
transmission and reflection coefficients between Bloch states on
either side of the interface. The details of WFM can be found in
Ref.[\onlinecite{master}].

The interfacial distortion was calculated based on Vienna
\emph{ab.initio} simulation package (VASP)\cite{VASP} with
ultrasoft pseudopotentials\cite{USPP} and generalized gradient
approximation\cite{GGA_PW} for exchange correlation energy. Common
lateral lattice constant used is the average value of the lattice
constant of both metals obtained in VASP, \emph{viz.}
$a_{cu}=a_{co}=3.595$\AA. The distortion was limited along [001]
direction. For Cu$|$Co interface, the modelling is based on the
supercell structure which consists of 6ML $fcc$ Co(001) and 6ML
Cu(001) with 12\AA vacuum inserted between them. The three layers
closed to vacuum at Co side and the three layers closed to vacuum
at Cu side are fixed. The in-plane lattice constant is fixed to
the common lattice constant. Energy cut-off is set to 400 eV and
$k$ grid is 19$\times$19$\times$1. For a given supercell the
structure is relaxed to reach the minimum energy. The cell size is
then varied, and a curve of structure \emph{vs} energy can be
obtained. The optimized structure with the lowest total energy is
deduced from this curve. For the surface Vac$|$Cu, the supercell
structure consists of 6ML Cu(001) and 12\AA  vacuum. Only the
three Cu layers closed to vacuum are allowed to move. The
optimized structure of Cu surface is obtained when the supercell
reaches it minimum energy after relaxation.

Due to work function mismatch, the distance between Cu and Co
layers is stretched to 1.86\AA. At Cu side the distance between
the neighboring layers is stretched by 3\%. At Co side, the
distance between the topmost two layers is contracted by 5\% and
the distance between the second and third layers is contracted by
1\%. At Cu surface, the distance between the topmost two surface
layers is contracted by 1\%. With the atomic structure obtained
from VASP as the input into the calculations of electronic
structure and transport. The volume of atomic sphere at interface
is a little different from that in bulk and selected to satisfy
the principle of full space filling. The unrelaxed part of the two
leads of interface are concatenated with bulk, where electronic
structure of bulk has been obtained independently and the
self-consistent potential of interface are treated as the
"embedding potential" of the bulk system with aid of
SGF\cite{master}.  We found the relaxation of Co side is crucial
to understand the spin dependent electron reflection correctly.

The spin dependent normal reflection with normal incidence in
Vac$|$Cu$|$Co(001) system is calculated. As there is only one
channel of $\triangle_{1}$ symmetry in copper at the $\Gamma$
point in the 2-dimensional Brillouin zone (2D BZ), the total
reflectivity back to the vacuum for each spin is
\begin{equation}\label{eq-1}
R=\frac{r^2_{B}+e^{-4d/\lambda}r^2_{C}-2e^{-2d/\lambda}r_{B}r_{C}cos(2k^{e}d-\phi_{B}-\phi_{C})}{1+e^{-4d/\lambda}r^2_{B}r^2_{C}-2e^{-2d/\lambda}r_{B}r_{C}cos(2k^{e}d-\phi_{B}-\phi_{C})}
\end{equation}
where $k^{e}\equiv k_{BZ}-k$, $k_{BZ}$ is the BZ wave vector, $k$
is the electron momentum vector. $r_{B}exp(i\phi_{B})$ and
$r_{C}exp(i\phi_{C})$ are the reflection coefficients from Cu side
at the Cu$|$Vac and Cu$|$Co, and $d$ is the thickness of Cu film.
Due to the inelastic scattering and the impurity scattering in Cu
film, a mean free path (MFP) $\lambda$ has to be incorporated,
where MFP in the energy scale is assumed to be uniform to minimize
our adjustable parameter in spite of possible energy dependence.
By this way, only elastic reflection are taken into account and no
inelastic reflection will be considered. The reflectivity
asymmetry is defined as $(R_{p}-R_{ap})/(R_{p}+R_{ap})$ with
$R_{p}$($R_{ap}$) denoting total reflectivity with incident spin
parallel to majority(minority) spin in Co.

To justify the theoretical calculation, we made the comparison
between our calculation and the experimental data quantitatively.
The experimental electron reflection spectrum for the electron
energy higher than the vacuum level was obtained by the
spin-polarized low-energy electron microscopy (SPLEEM)
measurement. The experiment was performed at Lawrence Berkeley
national laboratory.  The detailed experimental description can be
found in Ref.[\onlinecite{Co_Cu_YZWu}]. In the SPLEEM, a spin
polarized electron beam is directed at the sample surface at
normal incidence and the spin-dependent electron reflectivity can
be measured simultaneously during the Cu growth on 5ML
Co$|$Cu(001). As shown in Eq.(\ref{eq-1}), the maximum electron
reflectivity takes place at the interference condition of
$2k^{e}d-\phi_{B}-\phi_{C}=2\pi n$, where n=integer. So by
measuring the thickness dependent electron reflectivity at certain
electron energy, the $k^{e}$ and  the total phase gain
$\phi_{B}+\phi_{C}$ can be derived experimentally\cite{R. Zdyb}.
However, the phase gain calculated using this method has larger
error bar, which is larger than the PGD.  To obtain the PGD
precisely, we can determine the peak position difference $\Delta
d$ in the reflection spectrum of spin-up and spin-down electron,
then the PGD can be calculated by the formula $\Delta
\phi=2k^{e}\Delta d$.

The QW states for minority spin below the Fermi surface were
measured by the angular resolved Photoemission at beamline 7.0.1.2
of the Advanced Light Source at the Lawrence Berkeley National
Laboratory, and the experimental details can be found in
Ref.[\onlinecite{Cu_Co_ab_initio_phase}] and
Ref.[\onlinecite{Wu_2002}]. The electron density oscillates with
the Cu thickness, the periods will be determined by the $k^{e}$,
and the phase can be obtained by measuring the peak position. Due
to the high quality of our sample, the Cu thickness can be
determined precisely, so the phase gain for minority spin can be
obtained with high accuracy.

\begin{figure}
\includegraphics[width=7cm, bb=9 9 204 250]{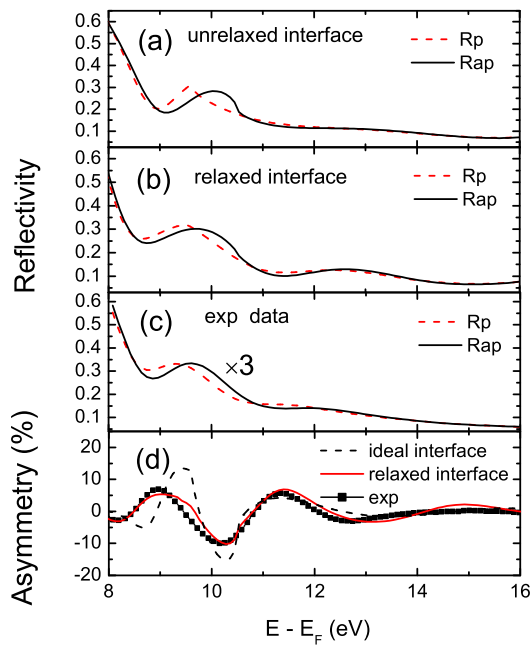}
\caption{\label{fig:24}(color online) The calculated spin
dependent electron reflectivity spectra with (a) unrelaxed
interface, (b) relaxed interface and (c) the experimental spectra
for the system of Vac$|$Cu(7ML)$|$Co(001). (d) The calculated
reflection asymmetry for the spectra shown in (a)-(c). The
experimental spectra have been multiplied by a factor 3 for better
comparison. The mean free path used in the calculation is 7ML. }
\end{figure}

\section{Numerical results and Discussion}

To identify the effect of the distorted lattice, we calculated the
energy dependence of reflectivity and reflectivity asymmetry for
Vac$|$Cu(7ML)$|$Co(001) system with unrelaxed and relaxed
interfaces shown in Fig.\ref{fig:24}, where the unrelaxed
interface is assumed to be of idea \emph{fcc} lattice. The
amplitude of theoretical reflectivity is not very sensitive to the
interfacial distortion. The amplitude obtained in experiment is
almost 3 factor less than theoretical results. Due to the
introduction of MFP in the Cu film, the amplitude is dominated by
the Vac$|$Cu surface. As the measurement of reflectivity is during
the growth of sample, the discrepancy mainly results from the
impurities and inelastic scattering at Cu surface. However, the
asymmetry spectra is more sensitive to the interfacial distortion,
since only the electron reflection at Cu$|$Co interface is spin
dependent. By quantitatively comparing our \emph{ab. initio}
results and experimental data as shown in Fig.\ref{fig:24} (c), we
can draw a conclusion that the interfacial distortion is very
crucial to obtain the correct spin-dependent electron reflection
spectrum. Here, we found that $\lambda$=7ML$\pm$1ML in Cu from 8eV
to 16eV gives the best reproduction of experimental data.
\begin{figure}
\includegraphics[width=7cm, bb=12 9 198 243]{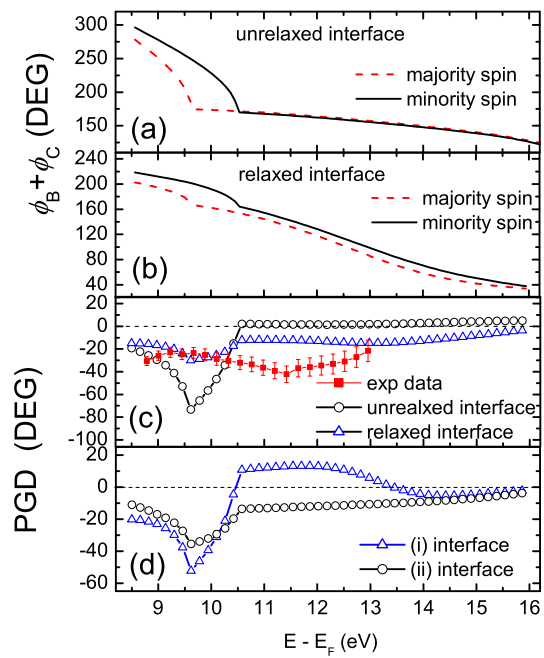}
\caption{\label{fig:20}(color online) (a) Total reflection phase
$\phi_{B}+\phi_{C}$ of the unrelaxed interfaces of Cu$|$Co and
Cu$|$Vac. (b) Total reflection phase $\phi_{B}+\phi_{C}$ of the
relaxed interfaces of Cu$|$Co and Cu$|$Vac. (c) Comparison of PGD
of unrelaxed, relaxed interface with experimental data. (d) PGD of
the two artificial interfaces discussed in the text. }
\end{figure}

The effect due to the distortion of interface is more obvious for
the phase gain at interface. In (a) and (b) of Fig.~\ref{fig:20},
the summation of the reflection phases of the two interfaces are
shown for unrelaxed structure and relaxed structure respectively,
where the shape of the spectra for two spins are similar to each
other and the odd points of reflection phases appear at the band
edge for each spin. The PGD spectra for the unrelaxed and relaxed
Cu$|$Co interfaces are shown in Fig.~\ref{fig:20} (c) with
experimental data. Sign of PGD have been reversed from positive
for the unrelaxed interface to negative for the relaxed interface,
which is consistent with experimental results. Note that LDA fails
to produce the image-like potential out of Cu
surface\cite{image_LDA_Lang}. With referring to the empirical
model\cite{image_Smith}, we have estimated the effect due to the
image states on the reflection phase at Vac$|$Cu by shifting the
potential at vacuum side to reproduce the surface
potential\cite{image_correction}. For energy far higher ($>$ 8eV)
than Fermi level, the effects due to the image states are small.

To clarify the source of PGD for the relaxed interface, we have
artificially constructed two interfaces (i) and (ii). In (i), the
structure at Co side of the relaxed Cu$|$Co interface was replaced
by the ideal {\it fcc} lattice. In (ii), the structure at Cu side
of the relaxed Cu$|$Co interface was replaced by the idea {\it
fcc} lattice. As shown in of Fig.~\ref{fig:20} (d), PGD of (i) is
very similar to the spectrum of the completely relaxed interface,
which indicates that Co side of the relaxed structure is the
decisive factor for the PGD of the relaxed Cu$|$Co interface. The
penetration of wave into the Co side will induce the spin
dependent phase accumulation and the distortion of this part could
affect greatly the PGD. The main difference of PGD between
experimental data and our theoretical results is the general shape
of the energy dependence. Two odd points exist in the theoretical
results, however, there is no such clear point in experimental
data at corresponding energy. Such difference can not be
attributed to the uncertainty of interfacial structure model, as
the odd points are due to the band edges of the two spin
species(see Fig.~\ref{fig:20} (a)\&(b)) and the interfacial
structure couldn't smear the mismatch of band structures of Cu and
Co. The possible contribution to the difference may come from the
electron-phonon (ep) coupling at interface which is not
incorporated in our calculation yet. The ep coupling has been
found could be important for the phase shift at metallic
interface\cite{E_ph_Luh02}. With the including of phonon, the
energy dependence of reflection phase is dominated not only by the
electron band structure but also by the phonon spectrum. It is
expected that the difference of the energy dependence of PGD may
be compensated by the mixing of electron band structure and phonon
spectrum.

\begin{figure}
\includegraphics[width=7cm, bb=17 16 281 221]{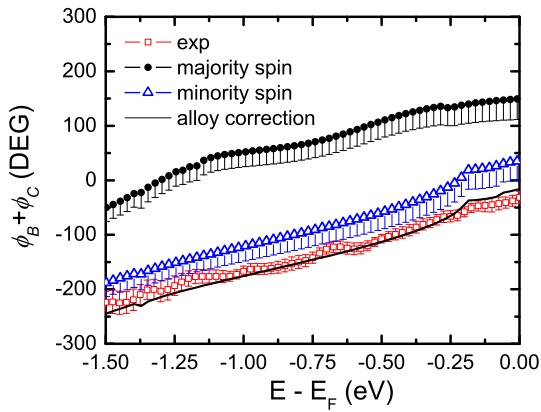}
\caption{\label{fig:25}(color online) The calculated reflection
phase $\phi_{B}+\phi_{C}$ of the majority spin(solid circle) and
minority spin(open triangle) with the experimentally measured
phase(open square). The solid line indicates the lower limit
(position of error bar) of minority spin after taking interfacial
alloy into account as discussed in text.}
\end{figure}

In experiment, one monolayer alloy could be located at interface.
It was found that variation of the PGD due to 1ML substitutional
alloy Cu$_{x}$Co$_{1-x}$($0<x<1$) inserted between Cu and Co is
small($< 2$ DEG), which is consistent with above discussion as 1
ML alloy only slightly changes the structure at Co side. However,
interfacial alloy might induce some extra phase gain when
electrons are reflected from the interface, which could change
both energy and Cu film thickness dependent spectra of
reflectivity.

Up to now, the phase discussed is above Fermi level. The phase
near Fermi level may be more meaningful as most electronic
transport experiments are carried out around this energy.
Experimentally, it is very hard to separate the contributions of
majority and minority spin at Fermi level. However, it is well
known that only the electrons with minority spin can form QW
states at Fermi level in Cu$|$Co(001) system. So the phase gain of
minority spin can be obtained experimentally. Fig.~\ref{fig:25}
shows our theoretical calculation\cite{Band_shift} and the
experimental data. As we discussed, the image states may have the
effect on the phase gain for the electron lower than Fermi level.
The including of image states shifts down the total phase up to 50
DEG in this energy scale as shown by the uncertainty bar. Such
large correction is important when comparing with experimental
data. Our theoretical results give the good reproduction of the
slop of total phase. The small discrepancy in magnitude can be
roughly compensated by replacing the layer of Cu at interface with
1ML Cu$_{0.3}$Co$_{0.7}$ alloy, which could further shift down the
phase spectra as shown by the solid line in Fig.~\ref{fig:25}. The
total phase of majority spin is also present, which is far from
the experimentally measured one.

For $\Gamma$ point in 2D BZ, we have shown that the relaxation of
atomic structure could greatly affect the spin transport. But,
this is not enough to understand the spin transport. To discuss
the effect of interfacial distortion on the quantities such as
interface resistance or mixing conductance, we should extend our
calculation to the whole 2D BZ at Fermi level. The results are
shown in Tab.~\ref{tab:table1}.
\begin{table}
\caption{\label{tab:table1} Interface resistance $SR$ in units of
$f\Omega m^{2}$, the mixing conductance $G_{mix}$ in units of
$10^{15}\Omega^{-1} m^{-2}$. }
\begin{ruledtabular}
\begin{tabular}{ccccc}
    system & $R_{\uparrow}$ & $R_{\downarrow}$ & $Re(G_{mix})$ & $Im(G_{mix})$ \\
  \hline
  unrelaxed CuCo (001) & 0.303 & 1.91\footnotemark[1] & 0.556 & -0.021 \\
  relaxed CuCo (001) & 0.304 & 2.42 & 0.547 & -0.021 \\
  relaxed Cu$_{50}$Co$_{50}$ (001) & 0.306 & 1.53& 0.531 &-0.026 \\
  unrelaxed AgFe (001) & 1.00 & 6.79 & 0.445 & -0.003 \\
  relaxed AgFe (001) & 1.04 & 6.10 & 0.444 & -0.0006 \\
  relaxed Ag$_{50}$Fe$_{50}$ (001) & 1.02 & 2.59 & 0.441 & 0.0002 \\
\end{tabular}
\end{ruledtabular}
\footnotetext[1]{The slight difference of $SR$ between present
results and those in Ref.[\onlinecite{Al_Ag}] comes from different
atomic potentials. Therein, $spd$ basis is used.}
\end{table} For Cu$|$Co interface,
interface resistance of majority spin is less affected, however,
minority spin is well affected by structure relaxation. For
majority spin, $d$ bands in Cu and Co are fully filled and those
states at Fermi level are from $s$ bands which are essentially
free electron like states. The interface could be equivalent to a
very low potential step for those states which have the same
symmetry in both materials. Interfacial distortion is just like
the variation of the potential profile, which is expected to have
little effect on the transport. For minority spin, $s$ band still
dominates in Cu, but $d$ bands in Co are only partly filled, which
lead to the multiple sheets of Fermi surface in Co. Interface
resistance should be attributed to the mismatch of band structure
in Cu and Co\cite{master}, e.g. mismatch of Fermi velocity and
mismatch of symmetry of wavefunction\cite{Al_Ag}. Interfacial
distortion could play important roles in those issues.
Furthermore, our results show that one layer of interfacial alloy
could enhance such effect. This can be understood as more states
of minority spin at Co side could be diffusively coupled to those
at Cu side with aid of the alloy\cite{master}.

However, $G_{mix}$ is not sensitive to interfacial distortion.
According to the definition of $G_{mix}$, there exists a term
$r_{\uparrow}^{nm}||r_{\downarrow}^{nm}|exp(i(\phi_{\uparrow}^{nm}-\phi_{\downarrow}^{nm}))$
with the summation of the 2D BZ. For the single point in 2D BZ
this term is sensitive to the interfacial distortion, however, the
summation could lead to the cancellation of this term among
different points in 2D BZ due to the phase factor
$\phi_{\uparrow}^{nm}-\phi_{\downarrow}^{nm}$. The real part of
$G_{mix}$ obtained in our calculation is very close to the value
of $\frac{e^{2}}{h}M$, which means strong cancellation did happen.
As a result, the sensitive issue of interface reflection phase
could be trivial one.

To confirm above argument, we also studied the Ag$|$Fe(001)
interface. The modelling of the interfacial relaxation is just
like that for Cu$|$Co interface, but note that in this case we
need to rotate the $bcc$ Fe(001) plane by 45 DEG to match the
$fcc$ Ag(001) plane. The lattice constant
$a_{Fe}=\frac{\sqrt{2}}{2}a_{Ag}=2.855$\AA. After relaxation the
distance between Fe and Ag layers is about 1.39 times that of
neighboring layers' distance in bulk Fe. The contraction of the
top two layers at Fe side is by 1\% and that at Ag side is less
than 1\%. Similar results were also obtained in this system. The
interface resistance of majority spin and the real part of
$G_{mix}$ are both less sensitive to the interfacial distortion
shown in Tab.~\ref{tab:table1}. The magnitude of the imaginary
part of $G_{mix}$ in this case greatly decreases for the relaxed
interface, but the imaginary part is too small compared with the
real part to contribute to spin transfer
torque~\cite{Brataas_circuit}.

\section{Conclusion}

The phase information obtained from experiments can be well
understood by our \emph{ab.initio} calculation in a large energy
scale, it was found that structural distortion at interface is
important for theoretical model to interpret experiment. The PGD
of two spins obtained in our calculation is comparable with
experimental results. At Fermi level, extra phase gain due to
image states becomes obvious. After integration of 2D BZ, we also
found that the interface resistance of minority spin can be well
affected by the interfacial distortion, however, the interface
resistance of majority spin and the PGD related quantity mixing
conductance are not sensitive to the interface structure.

We thank Professor Z. Q. Qiu for the helpful discussions. This
work is financially supported by NSF (Grant No. 10634070,
10604015, 10621063) and MOST (Grant No. 2006CB933000,
2006CB921300, 2006AA03Z402) of China.



\begin{references}
\bibitem{spdf}The potentials were obtained with $spdf$ basis set in this work instead of $spd$ basis set as in our previous work.
\bibitem{image_correction}Here, we only use the sphere potential to mimic the empirical model of image
state. The results are only used as an estimation of uncertainty
in our calculation.
\bibitem{Band_shift}Comparing with other theoretical and experimental
results[\onlinecite{band_Cu_Ag_Au, band_Cu_Co001,
Ferim_radii_Cu_Ag_Au}], our calculated $sp$ band which dominates
QW stats here is lower by around 0.30 eV. The DFT-LDA band
error[\onlinecite{LDA_error, Cu_Co_ab_initio_phase}] should
response to such discrepancy. The error is corrected by a rigid
shift of band.
\end{references}
\end{document}